\documentclass[twocolumn]{aastex63} 
\hypersetup{linkcolor=red,citecolor=blue,filecolor=cyan, urlcolor=black}
\usepackage{multirow}
\usepackage{amsmath}
\usepackage{amssymb}
\usepackage{balance}
\usepackage{lmodern}
\usepackage{changepage}
\usepackage{hyperref}
\usepackage{textcomp}

\received{June 5, 2026}
\revised{June 5, 2026}
\accepted{June 5, 2026}
\submitjournal{AAS Journals}

\shorttitle{Eppur binaria non \'e esclusa}
\shortauthors{Iwanek et al.}

\begin{document}

\title{Eppur binaria non \'e esclusa: Gaia astrometry does not disfavor a binary origin for Long Secondary Periods}

\correspondingauthor{Patryk Iwanek}
\email{piwanek@astrouw.edu.pl}

\author[0000-0002-6212-7221]{Patryk Iwanek}
\affiliation{Astronomical Observatory, University of Warsaw, Al. Ujazdowskie 4, 00-478 Warsaw, Poland}

\author[0000-0003-2244-1512]{Piotr A. Kołaczek-Szymański}
\affiliation{Astronomical Observatory, University of Warsaw, Al. Ujazdowskie 4, 00-478 Warsaw, Poland}
\affiliation{Astronomical Institute, University of Wrocław, Mikołaja Kopernika 11, 51-622 Wrocław, Poland}

\author[0000-0001-9439-604X]{Dorota M. Skowron}
\affiliation{Astronomical Observatory, University of Warsaw, Al. Ujazdowskie 4, 00-478 Warsaw, Poland}

\author[0000-0002-6495-0676]{Grzegorz Pojmański}
\affiliation{Astronomical Observatory, University of Warsaw, Al. Ujazdowskie 4, 00-478 Warsaw, Poland}

\author[0000-0002-7777-0842]{Igor Soszy{\'n}ski}
\affiliation{Astronomical Observatory, University of Warsaw, Al. Ujazdowskie 4, 00-478 Warsaw, Poland}

\begin{abstract}

\noindent We present an independent reassessment of the nearby long secondary period (LSP) stars analyzed by \citet{shariat2026}. By inspecting long-term All Sky Automated Survey (ASAS) light curves, together with Gaia Focused Product Release (Gaia FPR) radial velocity  (RV) curves, for 221 LSP candidates (out of 224) located within 1.5 kpc from the Sun, we find that less than half of the sample (103 objects, $\sim 47$\%) exhibit convincing LSP-like behavior. The remaining part of the sample could be more naturally interpreted as that of semi-regular variables (SRVs), characterized by irregular or multi-periodic pulsations. This indicates that the analyzed sample is significantly contaminated by non-LSP objects and therefore is not representative of the classical period--luminosity sequence-D population. Using the \texttt{gaiamock} tool to predict Gaia renormalized unit weight error (RUWE) values for binary systems, we show that even the nearest LSP stars do not have to exhibit elevated RUWE values as a~consequence of their binarity. We also argue that the binary-nature hypothesis for LSP stars does not lead to a discrepancy between the observed and expected distance--RUWE relation for these variables.

\end{abstract}

\keywords{
\href{https://astrothesaurus.org/uat/154}{Binary stars (154)};
\href{https://astrothesaurus.org/uat/1372}{Red giant stars (1372)};
\href{https://astrothesaurus.org/uat/1464}{Sky surveys (1464)};
\href{https://astrothesaurus.org/uat/498}{Exoplanets (498)}
}

\section{Introduction} \label{sec:introduction}

Binary stars have long served as one of the most useful laboratories in astrophysics. They provide an almost direct route to stellar masses, allow for calibration of evolutionary models, and when interaction between components becomes important, open evolutionary channels inaccessible to single stars. Mass transfer, common-envelope evolution, wind interaction, circumbinary dust, and tidal distortion are not marginal complications in stellar astrophysics; they are among its central mechanisms. Many classes of objects that once appeared to be peculiar or exceptional became physically intelligible only after their binary nature was recognized. In this sense, binarity is not only one possible explanation among many, but a recurrent warning that apparently single-star phenomena may carry the imprint of an unseen companion \citep[e.g.,][]{duchene2013, han2020, chen2024, boffin2025}.

Long secondary periods (LSPs) in luminous red giants belong to this category of long-standing puzzles. They manifest as photometric variations with periods of several hundred to several thousand days and amplitudes reaching up to 1 mag at the visual wavelengths, superimposed on shorter pulsation variability, with periods approximately one order of magnitude shorter, typical of red giant branch (RGB) and asymptotic giant branch (AGB) stars \citep[e.g.,][]{soszynski2021, iwanek2025}. In the period-luminosity diagram they form the well-known sequence D, distinct from the sequences associated with radial and non-radial pulsations \citep{wood1999, wood2000b, soszynski2007}. The LSP phenomenon has been recognized for nearly a century, beginning with the early reports of secondary periods in red variables by \citet{oconnell1933}, followed by broader observational studies by \citet{payne1954} and \citet{houk1963}. However, despite the enormous progress made thanks to data from large-scale time-domain photometric surveys \citep[e.g.,][]{wood1999, wood2000b, fraser2008, soszynski2009, soszynski2011, soszynski2013, pawlak2019, gaiafpr2023, lebzelter2023}, the physical origin of LSPs remains unclear.

The difficulty in solving the mystery of LSP stars is not due to their rarity. LSPs are common, occurring in roughly one third of the pulsating red giants. This phenomenon often persists over many cycles, shows characteristic light curve shapes\footnote{https://ogle.astrouw.edu.pl/atlas/LSPs.html} \citep{soszynski2014}, and is accompanied by variations in radial velocity (RV) with full amplitudes of a few kilometers per second \citep[][]{nicholls2009}. They are also linked to infrared excesses and dust-related variability, suggesting that the outer atmosphere or the wind of the giant participates in the phenomenon \citep{wood2009}. Therefore, any potential solution must explain not only the timescale but also the coherence, morphology, RVs, and circumstellar signatures.

Several mechanisms have been proposed to explain the LSP phenomenon. Non-radial pulsations \citep{wood2000a, wood2004, saio2015, takayama2020}, rotational modulation by large convective cells \citep{stothers2010}, and episodic dust formation \citep{wood2004, takayama2015} have all been invoked to account for parts of observed behavior. However, no single-star model has yet provided a fully satisfactory explanation of the complete set of observables. In particular, reproducing long periods while simultaneously explaining RVs, the morphology of the light curve, and infrared behavior remains challenging.

For this reason, binary interpretation has still remained one of the most plausible theories. In its simplest form, the LSP is associated with the orbital period of a low-mass companion embedded in, or interacting with, the extended atmosphere and wind of the red giant. The observed photometric modulation may then arise from variable obscuration by a dusty cloud, wake, or trailing structure associated with the companion, whereas the infrared behavior may reflect the thermal emission and eclipses of this dusty material \citep{wood2009, soszynski2021, goldberg2024, macleod2025, decin2025}. Recent mid-infrared observations have strengthened this picture: many LSP stars show secondary minima visible in the infrared, naturally interpreted as eclipses of warm dust associated with the companion \citep{soszynski2021}. Hydrodynamical calculations and studies of dusty asymmetries around evolved stars further show that even low-mass companions reshape the wind of a red giant and create non-axisymmetric dusty structures \citep{decin2025, danilovich2025}.

If the binary interpretation of the LSP is correct, the implications turn out to be far-reaching. LSPs would not represent an additional variability class among red giants, but a common manifestation of interaction between evolved stars and low-mass companions. The incidence of LSPs would then imply that such companions survive, migrate, accrete, or interact in other ways with red giant envelopes far more frequently than might be inferred from main-sequence companion statistics alone. The phenomenon would become relevant to the late evolution of planetary and substellar systems, the shaping of dusty winds, mass loss prescriptions on the RGB and AGB, and possibly the formation of more exotic post-interaction systems. A binary origin of LSPs would therefore turn sequence-D stars into a powerful diagnostic sample, allowing for tracing complex interactions between low-mass companions and the extended atmospheres of evolved stars, as well as their outcomes.

The binary scenario also faces some problems. Interpreting the observed RVs of LSP stars as a purely Keplerian motion often implies companion masses in the substellar or very low-mass stellar regime, at separations of the order of a few astronomical units \citep{nicholls2009}. This mass range overlaps with the so-called brown dwarf desert, where companions are relatively rare around solar-type main-sequence stars \citep{marcy2000, grether2006}. Reconciling this apparent rarity with the high occurrence rate of LSPs is one of the most important challenges for the binary model. However, this tension depends on how the companions formed and evolved. One cannot also exclude the possibility that the aforementioned tension is an artifact resulting from the Keplerian treatment of the RV curves, which is not necessarily correct. Their initial properties may have been altered by mass accretion, orbital migration, or interaction with the expanding envelope of giants \citep{soszynski2021, decin2025}. Thus, the brown dwarf desert is a serious constraint, but not necessarily a decisive argument against binarity.

Recently, \citet[][hereafter S26]{shariat2026} proposed a new test of binary interpretation for LSPs using Gaia astrometry. In the analysis, they used long-period variables (LPVs) from the Gaia Focused Product Release \citep[Gaia FPR;][]{gaiafpr2023}, combining multi-epoch RVs, photometry and astrometry. Using the catalog of LPVs from Gaia FPR, the authors identified 224 LSP candidates within 1.5 kpc of the Sun, based on the custom period and photometric amplitude criteria. Then, under the assumption that the observed RVs trace Keplerian orbital motion caused by low-mass companions, they predicted that nearby LSP variables should exhibit an elevated renormalized unit weight error (RUWE) in Gaia DR3, reflecting unmodeled astrometric motion of the photocenter. Since the observed RUWE distribution of their sample remains close to that expected for apparently single stars, the authors argue that Gaia astrometry disfavors low-mass stellar or substellar companions as the dominant origin of LSPs.

It should be clearly emphasized that this is an important and timely argument in the discussion about the origin of LSPs. Gaia astrometry offers an independent test of binarity, and RUWE has been found to be useful as a practical indicator of unresolved astrometric perturbations in many contexts \citep{belokurov2020, elbadry2021, stassun2021}. At the same time, RUWE is a summary statistic of the quality of a single-star astrometric solution, affected by the Gaia scanning law, brightness of a source, crowding, photometric variability and its period, photocenter motion, and, most importantly, the time span of the observations. Therefore, a low RUWE value \citep[usually below $1.4$; ][]{lindegren2018, fabricius2021} can be informative but does not automatically exclude the presence of a companion. Even more importantly, any population-level inference made based on RUWE is valid only as long as the purity of the studied sample can be considered satisfactory.

In this paper, we revisit the sample of nearby LSPs from the Gaia FPR analyzed by S26 and their RUWE simulations. The paper is organized as follows. In Section \ref{sec:rationale} we discuss the rationale of this paper and our approach. Section \ref{sec:ASAS} presents details on the data. In section \ref{sec:classification} we describe the classification procedure and data, which we publish within this paper, while in Section \ref{sec:samplecharacterization} we discuss the results of Gaia FRP LSP sample re-evaluation. Section \ref{sec:gaiamock-main} shows our approach to RUWE simulations and results. In Section \ref{sec:conclusions} we briefly discuss the results, while in Section \ref{sec:summary} we summarize the paper.

\section{Rationale and approach} \label{sec:rationale}

The main goal of this paper is to test whether the stars used in the analysis of S26 are bona fide LSP variables, and whether their Gaia DR3 RUWE values can be used to disfavor a binary origin of the LSP phenomenon.

Verification of the sample is necessary because the physical interpretation of the LSP phenomenon is critically dependent on the purity of the sample. An LSP is not defined solely by the presence of long-timescale variability but by a distinct secondary period superimposed on the shorter pulsation variability of a red giant; therefore, a clear difference between brightness amplitudes of both signals is required. Establishing this behavior requires photometric monitoring over a time long enough to cover at least several cycles of the LSP to separate it from irregular, multi-periodic, or slowly evolving semi-regular variability.

This point is particularly important for Gaia-selected LSP candidates. Unfortunately, for LSP stars with periods of hundreds of days (with a mean value of 1.6 years), Gaia photometry, spanning about 3 years, often does not provide enough well-sampled cycles to allow for a reliable and unambiguous classification.

The need for such validation is strengthened by the possible physical nature of the LSP. If LSPs are connected with enhanced stellar winds, dust formation, or companion-driven structures in the extended atmospheres of red giants, the photometric manifestation of the phenomenon may evolve over time. The LSP signal seen in the light curve may appear, weaken, disappear, or change its morphology as the circumstellar environment changes. Such an evolution should mainly affect the observed light curve properties, not necessarily the underlying binary configuration. If the phenomenon is driven by a companion, the companion should remain present even when the photometric LSP signature becomes less regular or less prominent. Therefore, RV and RUWE need not change in the same way as the light curve. Consequently, the classification of a star as an LSP can depend on the epoch and duration of the observations, and short or sparsely sampled time series data may blur the distinction between genuine sequence-D stars and semiregular variables (SRV) contaminants.

Our approach is therefore observationally conservative. We use the LSPs sample from the Gaia FPR analyzed by S26 as the starting point, but before interpreting its RUWE distribution, we independently reassess the variability nature of its members. 
For this purpose, we rely primarily on long-term {\textit V}-band light curves from the All Sky Automated Survey \citep[ASAS;][]{pojmanski1997, pojmanski2001, pojmanski2002}, which are many times longer than the typical time-span of the Gaia light curves.

In addition, we inspect the Gaia FPR RV curves as an independent diagnostic of long timescale variability and as a consistency check for variability potentially related to orbital motion. We do not use RVs as the sole criterion for LSP, but they provide useful complementary information.

After this photometric reassessment, we perform astrometric simulations analogous to those presented by S26. We first apply the simulation framework to the S26 sample of nearby LSPs from the Gaia FPR, to reproduce and test the reference result. Then, we repeat the analysis using a revised LSP sample, supplemented with already known and independently classified nearby bright LSP variables from \citet{iwanek2025}.

\section{ASAS data} \label{sec:ASAS}

ASAS is a long-term, ongoing observational project dedicated to monitoring the photometric variability of bright stars. It began in 1997 at the Las Campanas Observatory in Chile (operated by the Carnegie Institution of Washington). A few years later, the northern observing station was established at the Las Cumbres Observatory on Haleakala (Maui, Hawaii, USA). The commercial CCD cameras and telephoto lenses used in the project, data acquisition systems and reduction software, have been modified and improved numerous times over the project's nearly 30-year lifespan.

Southern data (up to declination $+28^\circ$) were initially collected using Apogee AP-10 2Kx2K CCD cameras, $V$~and~$I$ filters, and 200 mm f/2.8 
telephoto lenses. Until 2003, observations were made in the form of single exposures lasting 180 s, resulting in an overexposure of stars with a magnitude of $V < 8.5$. Later, data were collected in triple exposures each lasting 60 s, reducing the saturation limit to $V \sim 7.5$ mag at the cost of slightly increased readout noise. Since 'bleeding' of the overexposed pixels was also addressed, photometry of even brighter stars can be used as long as the high noise level ($\gtrsim 0.1$ mag) and zero-point uncertainty are acceptable.

The sensitivity limit of this instrumental setup was about $V~=~14.5$ mag with an accuracy of about 0.3 mag. The optimal range for searching for variable stars with small amplitudes was therefore limited to $V = 13$ mag.

In 2010, new cameras (FLI ProLine 16000 with 4Kx4K CCDs) and larger 
aperture lenses (200 mm f/2.0) were installed. At the same time, the 
observation scheme was changed: two exposures were taken -- one lasting 
180 s and the other 18 s, which allowed the saturation limit to be reduced to about 5.5-6 mag and the survey depth to be increased to about $V = 15.5$ mag.

In the north, from 2006 to 2017, two Apogee Ap-10 cameras, 
200mm/2.0 lenses with $V$ and $I$ filters, and single exposures of 180 s were used to observe stars with a declination greater than $-40^\circ$. 
Measurements of stars brighter than $V \sim 8.3$ mag are significantly 
affected by saturation.

The wide field of view of the ASAS cameras (approximately 10x10 degrees) 
and relatively large camera pixels limit the resolution to about 10-15 
arcsec per pixel. This causes stellar images to blend together, 
particularly in the Milky Way. In this situation, profile photometry or 
image subtraction methods should yield the best results. However, due to 
the strong variability of the stellar PSF profiles in the wide field of view caused by the optics and mechanical characteristics of the telescope drive, more consistent results were obtained using aperture photometry. Apertures with diameters ranging from 2 to 6 pixels were used, and the selection of the aperture should be made individually depending on the star's brightness, the stellar field density, and the position in the field of view.

Although the photometric zero point of each observed field is determined 
from the stars in the Tycho catalog to an accuracy of about 0.05 mag, the 
individual zero points for individual stars may be subject to much 
larger errors due to blending and vignetting.

More details on the instruments and the data acquisition and reduction pipe-line can be found in earlier publications, e.g. \citet{pojmanski1997, pojmanski2001, pojmanski2002}.

\begin{deluxetable*}{lccclccr}
\tablecaption{Variability classification of the LSPs sample from the Gaia FPR.} \label{tab:classification}
\tabletypesize{\scriptsize}
\tablewidth{0pt}
\tablehead{
\colhead{ASAS ID} &
\colhead{R.A.} &
\colhead{Decl.} &
\colhead{$P_1$} &
\colhead{Gaia DR3 ID} &
\colhead{RUWE} &
\colhead{$d$} &
\colhead{Class} \\
\colhead{} &
\colhead{(h:m:s)} &
\colhead{($^\circ$:$'$:$''$)} &
\colhead{(d)} &
\colhead{} &
\colhead{} &
\colhead{(pc)} &
\colhead{}
}

\startdata
000528-4253.2 & 00:05:28.00 & -42:53:14.61 & 536.40874 & 4994896061572855680 & 1.246 & 503 & LSP \\
001058-1834.4 & 00:10:57.93 & -18:34:23.13 & 411.67680 & 2366204140290386816 & 1.233 & 404 & LSP \\
002251+5156.4 & 00:22:51.43 & +51:56:25.11 & 34.65822 & 418924929598754304 & 1.108 & 1441 & SRV \\
\multicolumn{8}{c}{$\vdots$} \\
232523-5458.2 & 23:25:23.05 & -54:58:14.62 & 379.94065 & 6499365277922571904 & 1.075 & 1464 & LSP \\
233032+5543.7 & 23:30:31.54 & +55:43:39.42 & 50.33954 & 1997702234134342784 & 1.193 & 1133 & SRV \\
233923+4158.9 & 23:39:22.84 & +41:58:56.86 & 56.95452 & 1922291507583273216 & 1.115 & 1277 & SRV \\
\enddata
\tablecomments{Table is sorted in increasing right ascension order. Only the first and last three rows are shown. The full machine-readable table is available online through the URL: \url{https://www.astrouw.edu.pl/asas/?page=XXX}}
\end{deluxetable*}

\section{Classification of the LSPs from the Gaia FPR sample} \label{sec:classification}

We reconstructed the sample of nearby LSPs from Gaia FPR analyzed by S26 using the table \texttt{tlsp\_merged\_gaia\_edenhofer\_corr.fits} provided by the authors on Zenodo\footnote{\url{https://zenodo.org/records/19412352}}. We retained all sources with \texttt{dist\_pc\_used\_dust} $\leq 1500$ pc. This cut returns 224 stars, exactly corresponding to the sample described by S26.

\begin{figure}
    \epsscale{1.2}
    \plotone{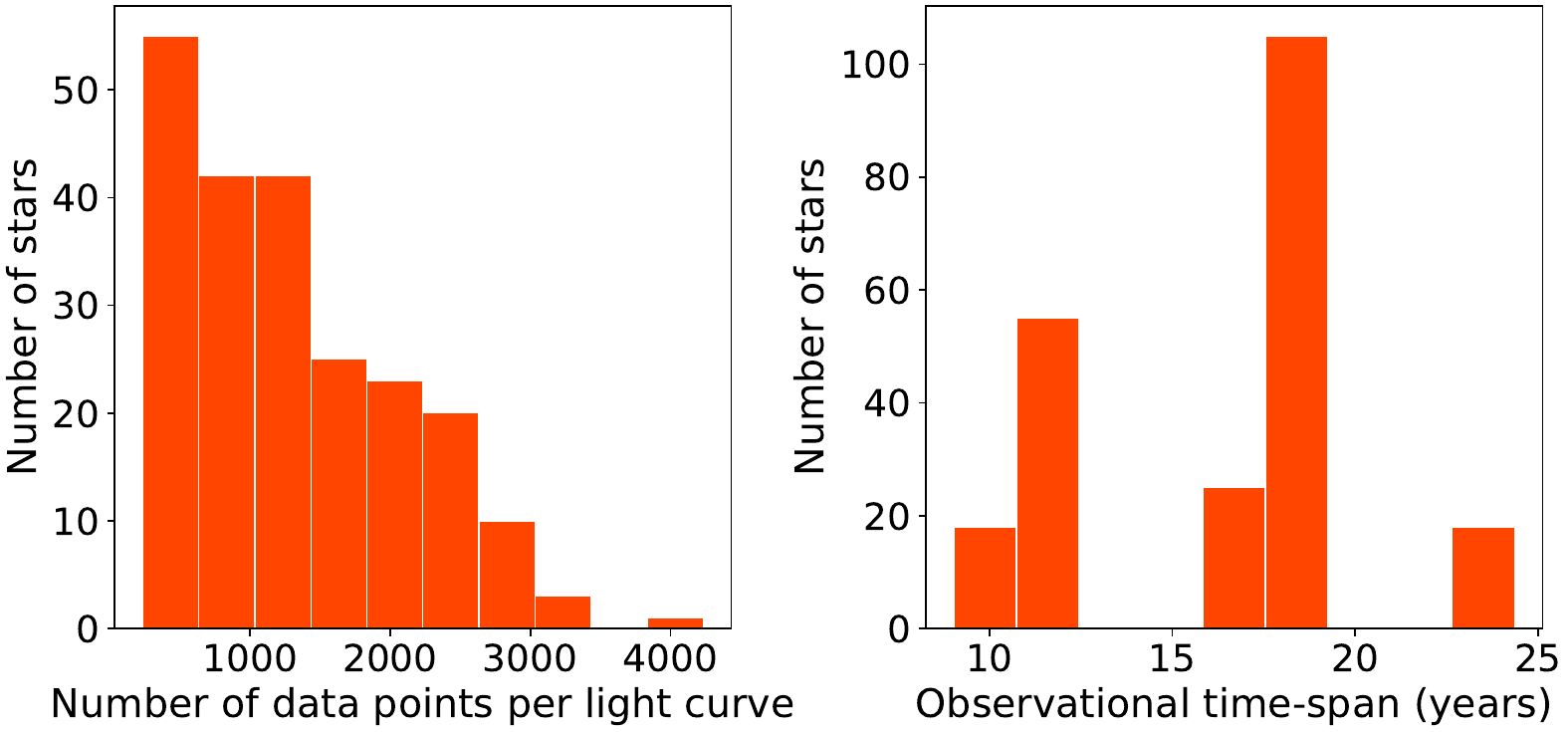}
    \caption{Summary of the photometric sampling for ASAS light curves of LSP candidates from Gaia FPR sample. {\textit{Left panel}}: the distributions of the number of epochs per light curve. {\textit{Right panel}}: the observational time-span in years per star.}
    \label{fig:asas_overview}
\end{figure}

We were able to extract ASAS photometry for 222 of the 224 objects. No ASAS light curves were found for two sources: Gaia DR3 533176454152832512 and Gaia DR3 2007994281452374784. For Gaia DR3 284016192897863808, only two photometric epochs were available, making meaningful variability classification impossible. Therefore, our final sample consists of 221 stars. For each object, ASAS provides photometry measured with five different aperture sizes. We therefore inspected all apertures for each light curve and, for the subsequent analysis, selected the aperture that yielded the smallest dispersion in the measured magnitudes. The basic characteristics of the ASAS light curves, including the number of epochs per light curve and the time-span of the observations, are presented in Figure~\ref{fig:asas_overview}.

For all 221 objects, we also retrieved the RV time series available in the Gaia FPR \citep{gaiafpr2023}. These data were used as complementary information to the ASAS light curves, particularly in cases where the photometric classification was ambiguous or the characteristic LSP morphology was not clearly visible in the ASAS data.

In the next step, we search for periodic signals using \texttt{fnpeaks}\footnote{\url{http://helas.astro.uni.wroc.pl/deliverables.php?active=fnpeaks}} software. We searched the frequency space from $0.0005$ to $0.1$ d$^{-1}$ with a frequency resolution of $10^{-5}$ d$^{-1}$, which corresponds to periods from $10$ to $2000$ days. We selected the period with the highest signal-to-noise ratio as the dominant one.

Finally, we carried out the most time-consuming part of the classification procedure: the visual inspection of each individual object. For every star, we examined the unfolded ASAS light curve, phase-folded ASAS light curve, and phase-folded Gaia FPR RV with the identified period. During this process, if the initially found period did not phase the light curve satisfactory, it was manually revised to obtain the most coherent representation of the long-term variability.

For the Gaia FPR RV, we additionally fitted a simple Keplerian orbit model. In these fits, the orbital period was fixed to the period identified in the ASAS light curve, while the free parameters were: the semi-amplitude of radial velocity of the primary component $K_1$, the time of periastron passage $T_0$, the systematic radial velocity $\gamma$, the argument of periastron $\omega$ and the eccentricity $e$. We emphasize that these fits were not intended to provide precise orbital solutions. Their purpose was only to aid the visual inspection of the phase-folded RVs and to assess whether the RV variability was broadly coherent with the photometric period.

To avoid biasing the reassessment against the LSP interpretation, we adopt an intentionally inclusive classification procedure. This was motivated by the possibly non-stationary photometric manifestation of LSPs: the signal may appear, weaken, disappear, or change morphology, potentially due to evolving configurations of circumstellar dust, the red giant, and its putative companion. We therefore did not require an unambiguous LSP morphology in the ASAS light curve in all cases. Candidates with weak or ambiguous photometric evicenve were retained as LSP when their Gaia FPR RV curves showed coherent periodic variability consistent with the LSP period, as dynamical and astrometric signatures of binarity should remain broadly stable even when the photometric footprint is weakened or temporarily absent.

Thus, our revised classification is conservative in the sense that it tends to preserve, rather than reject, the LSP nature of borderline objects. Objects were removed from the S26 LSP sample only when both the light curve morphology and the RV did not support the LSP interpretation.

This procedure led to a classification of all 221 objects. We identified 103 stars as bona fide LSP variables ($\sim 47\%$ of the original sample of nearby LSP candidates from the Gaia FPR). The remaining 118 stars were classified as SRVs. Thus, more than half of the sample used in the analysis of S26 does not show convincing LSP behavior.

The final classification, together with the ASAS identifier, the Gaia DR3 source ID, the dominant period ($P_1$), RUWE, and the distances adopted from S26, is presented in Table~\ref{tab:classification}. For confirmed LSP variables, $P_1$ refers to the LSP. The full machine-readable version of this table, as well as the ASAS {\textit V}-band light curves used in the classification, are made publicly available at:

\url{https://www.astrouw.edu.pl/asas/?page=XXX}

\section{Gaia FPR LSPs sample characterization} \label{sec:samplecharacterization}

\begin{figure*}
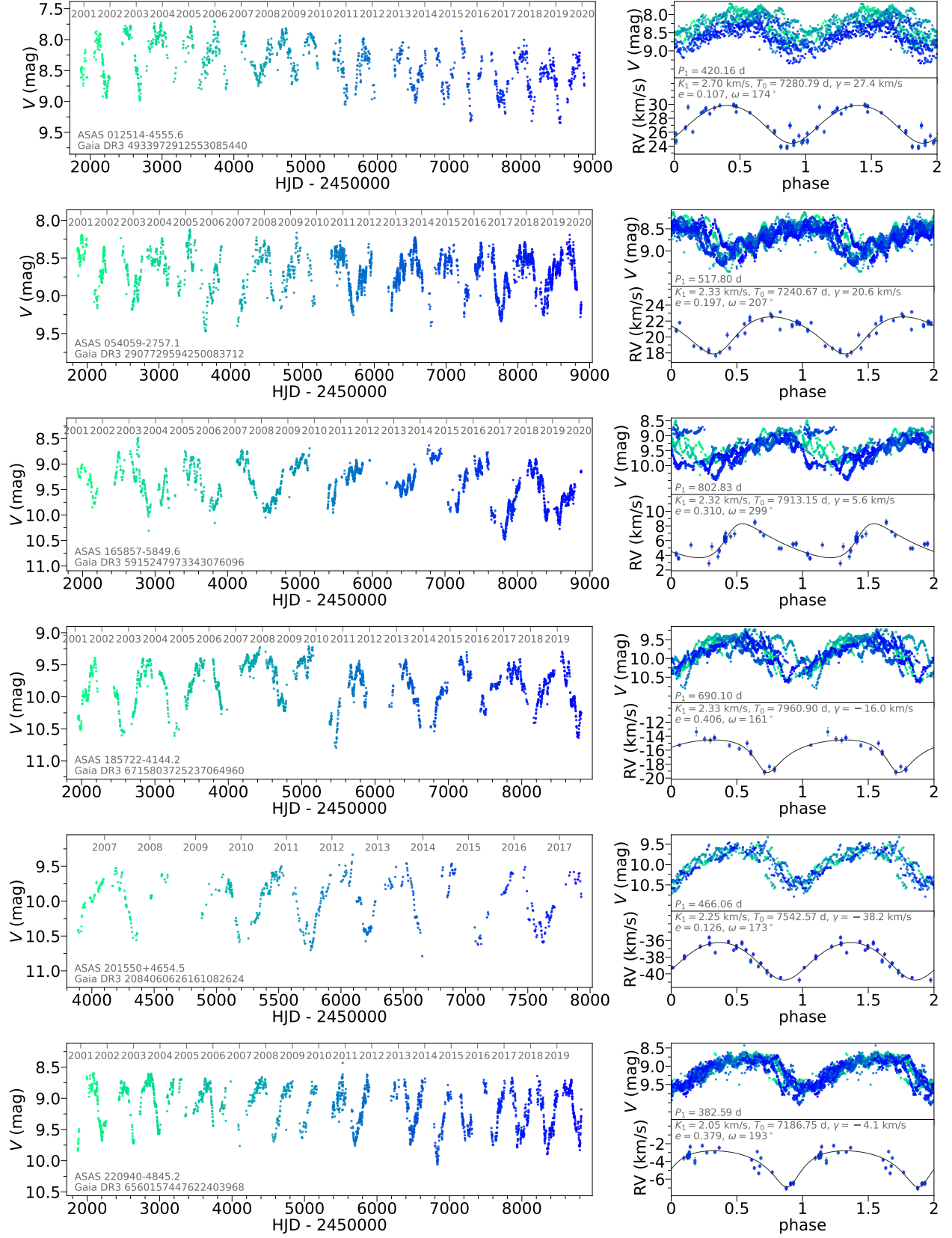

    \epsscale{1.1}
    \plotone{figures/LSP/012514-4555.6.pdf}
    \plotone{figures/LSP/054059-2757.1.pdf}
    \plotone{figures/LSP/165857-5849.6.pdf}
    \plotone{figures/LSP/185722-4144.2.pdf}
    \plotone{figures/LSP/201550+4654.5.pdf}
    \plotone{figures/LSP/220940-4845.2.pdf}
    \caption{Six examples of genuine LSP variables from S26 sample. Each LSP star is represented by three panels. {\textit{Left panel}}: unfolded ASAS light curve. At the top of this panel we marked years of observations, while inside the panel we plot ASAS ID and Gaia DR3 ID. HJD epochs are color-coded. {\textit{Top right panel}}: phase-folded light curve with the LSP ($P_1$) indicated inside the plot. Epochs are color-coded as in the top panel. {\textit{Bottom right panel}}: phase-folded RV curve with the LSP ($P_1$). Epochs are color-coded as in the top panel, while fitted Keplerian orbit model, with the parameters indicated inside this plot, is presented by black, solid line.}
    \label{fig:lsp1}
\end{figure*}

Our classification procedure shows that the nearby Gaia FPR LSP sample analyzed by S26 is highly heterogeneous when both photometric and RV information are considered. Although all 221 objects were originally treated as LSP variables, only 103 stars ($\sim 47\%$) show evidence consistent with genuine LSP behavior. In most cases, these objects exhibit the characteristic light curve morphology expected for LSP stars: a clearly visible LSP with an amplitude typically larger than that of the shorter pulsation period. The long-period modulation is recurrent and sufficiently coherent to be distinguished from semi-regular variability. Representative examples of our confirmed LSP stars, with clear evidence of LSP visible in both light curves and RVs are shown in Figure~\ref{fig:lsp1}.

\begin{figure*}
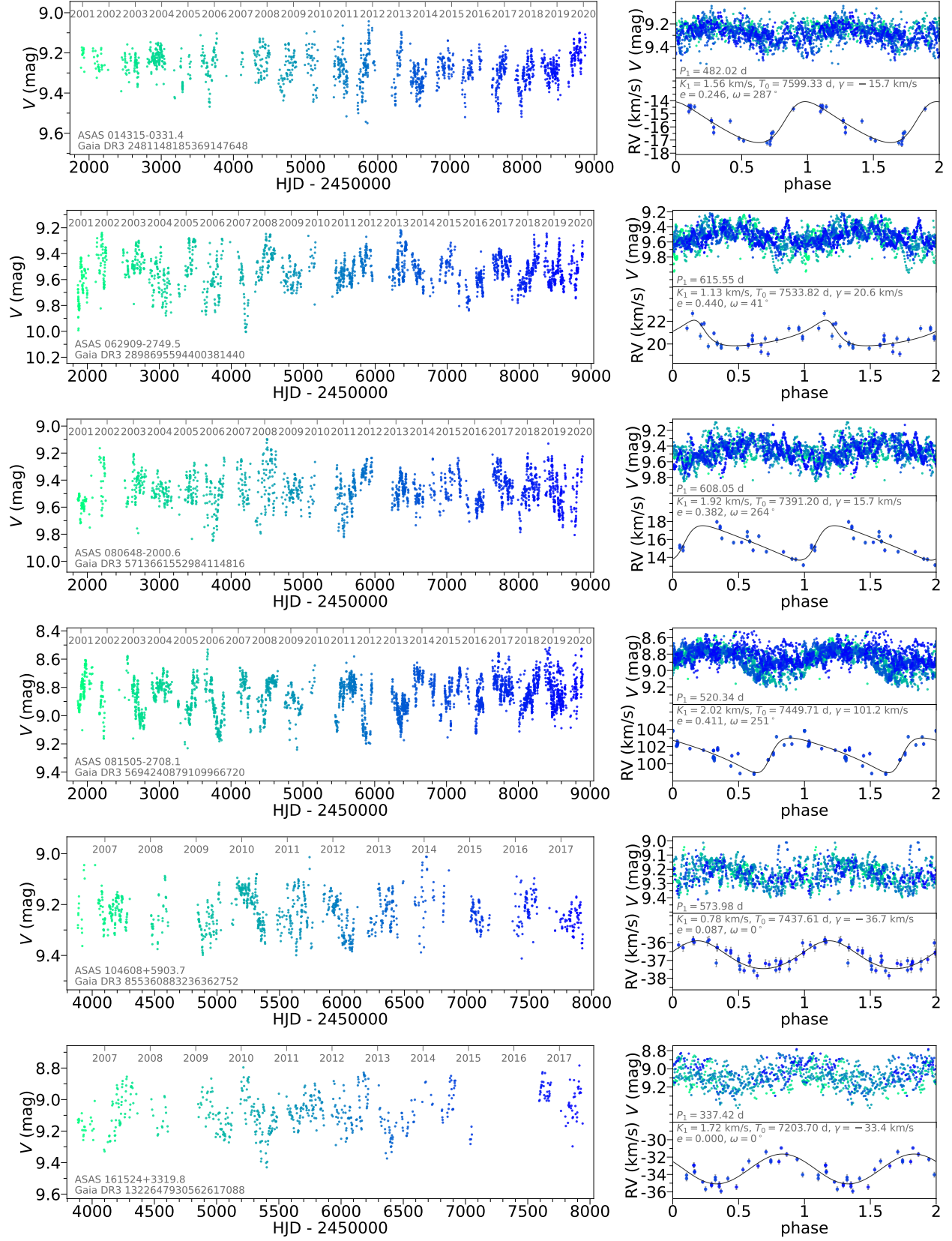

    \epsscale{1.1}
    \plotone{figures/LSP/014315-0331.4.pdf}
    \plotone{figures/LSP/062909-2749.5.pdf}
    \plotone{figures/LSP/080648-2000.6.pdf}
    \plotone{figures/LSP/081505-2708.1.pdf}
    \plotone{figures/LSP/104608+5903.7.pdf}
    \plotone{figures/LSP/161524+3319.8.pdf}
    \caption{Same as Figure \ref{fig:lsp1}, but six other examples of LSP are presented, for which the LSP morphology is either weaker or not as clearly separated from the shorter pulsation variability as in Figure~\ref{fig:lsp1}.}
    \label{fig:lsp2}
\end{figure*}

In Figure \ref{fig:lsp2} we show examples of objects that we also classified as LSP stars, although their photometric signatures are less clear than in the most straightforward cases. In these stars, the ASAS light curves show a long-timescale modulation, but the LSP morphology is either weaker or not as clearly separated from the shorter pulsation variability. However, Gaia FPR RVs phase with the same long period identified in the light curve, providing additional support for the LSP interpretation. These examples illustrate how RVs can help to retain objects whose photometric behavior alone would be more ambiguous.

\begin{figure*}
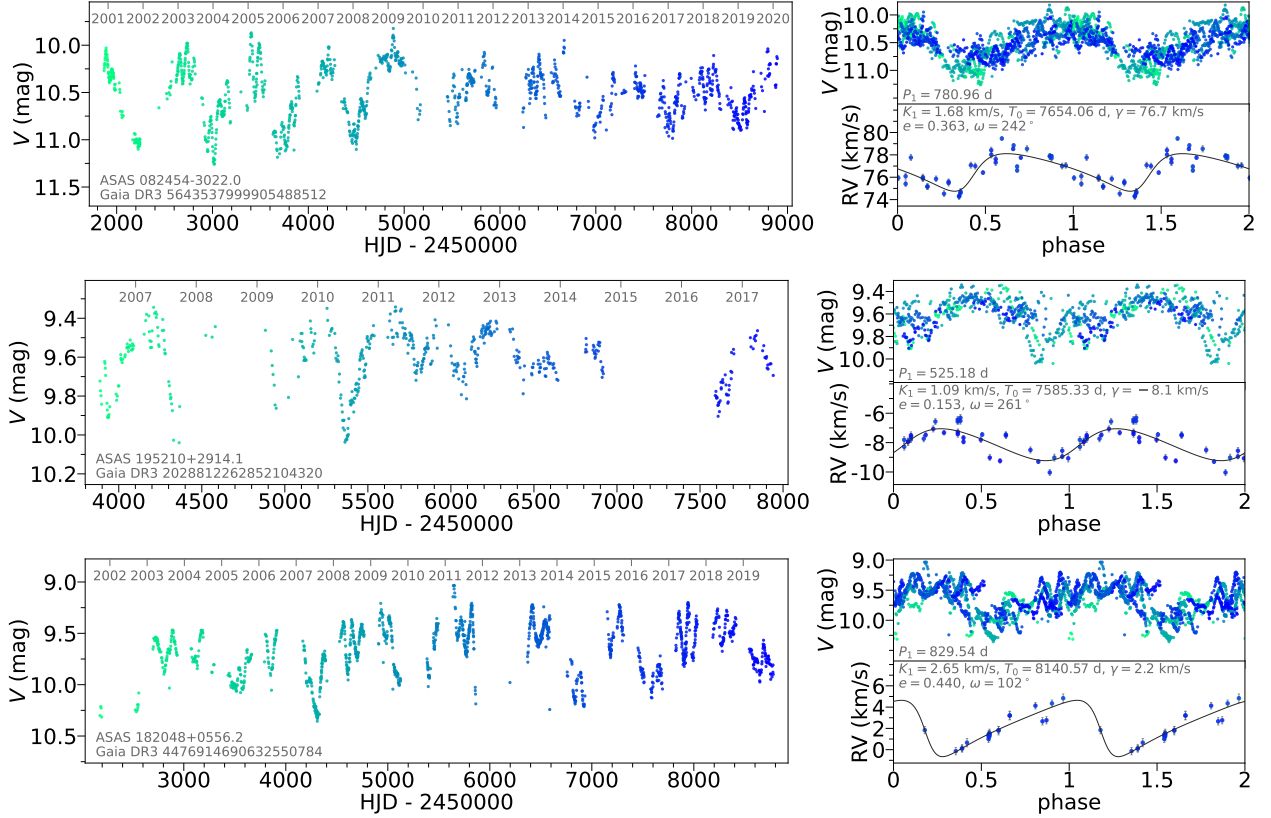

    \epsscale{1.1}
    \plotone{figures/LSP/082454-3022.0.pdf}
    \plotone{figures/LSP/195210+2914.1.pdf}
    \plotone{figures/LSP/182048+0556.2.pdf}
    \caption{Same as Figure \ref{fig:lsp1}, but here we present three examples of LSP stars in which a well-visible LSP becomes progressively weaker or disappears from the light curve.}
    \label{fig:lspweakenes}
\end{figure*}

\begin{figure*}
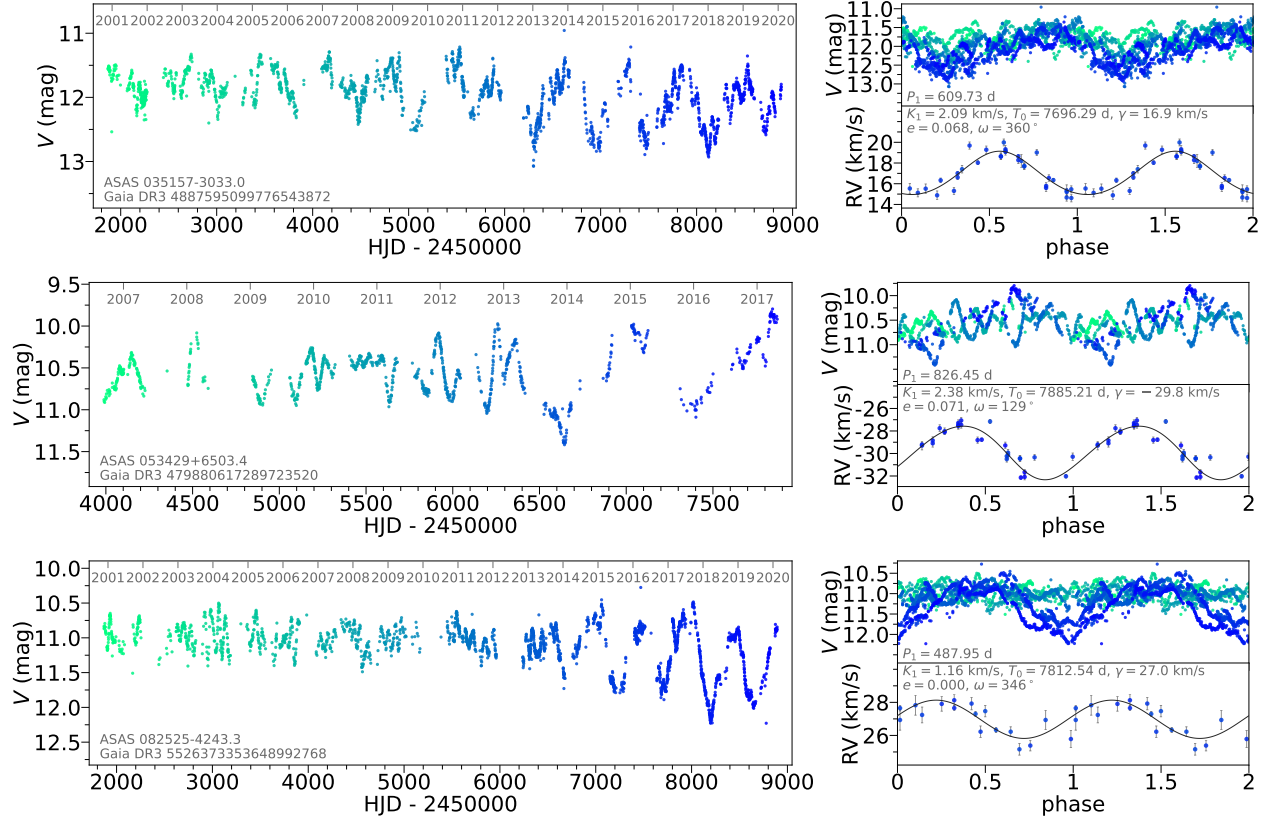

    \epsscale{1.1}
    \plotone{figures/LSP/035157-3033.0.pdf}
    \plotone{figures/LSP/053429+6503.4.pdf}
    \plotone{figures/LSP/082525-4243.3.pdf}
    \caption{Same as Figure \ref{fig:lsp1}, but here we present three examples of LSP stars in which a well-visible LSP becomes progressively stronger or appears in the light curve after being initially absent.}
    \label{fig:lspappearance}
\end{figure*}

The long ASAS light curves also reveal particularly interesting LSP objects whose behavior changes with time. In Figure \ref{fig:lspweakenes}, we show examples in which a well-visible LSP becomes progressively weaker or disappears from the light curve. Importantly, for these objects, the Gaia FPR RVs were obtained during epochs when the photometric LSP signatures were weaker or no longer clearly visible. Nevertheless, the RV variations remain phase-coherent with the long-period photometric signal identified in the light curves. This suggests that the disappearance of the LSP from the photometric data does not necessarily imply the disappearance of the underlying dynamical configuration that may be responsible for this phenomenon. On the other hand, in Figure \ref{fig:lspappearance}, we present objects that initially show predominantly semi-regular variability, while a distinct long period modulation appears in the ASAS time series. These cases may be especially important for understanding the LSP phenomenon because they suggest that the photometric manifestation of LSP variability is not necessarily permanent but may appear, weaken, disappear, or change its morphology on timescales comparable to the duration of modern large-scale sky surveys.

The existence of such objects indicates that the classification of red giant variables is not always time-independent. The same star may appear as an SRV when observed during one interval and as an LSP when observed during another, depending on whether the long-period modulation is present and detectable in the light curve at that time. This has important implications for samples constructed from relatively short or sparsely sampled photometric data. In such cases, the observed variability class may reflect not only the intrinsic nature of the star but also the particular temporal window covered by the observations. 

\begin{figure*}
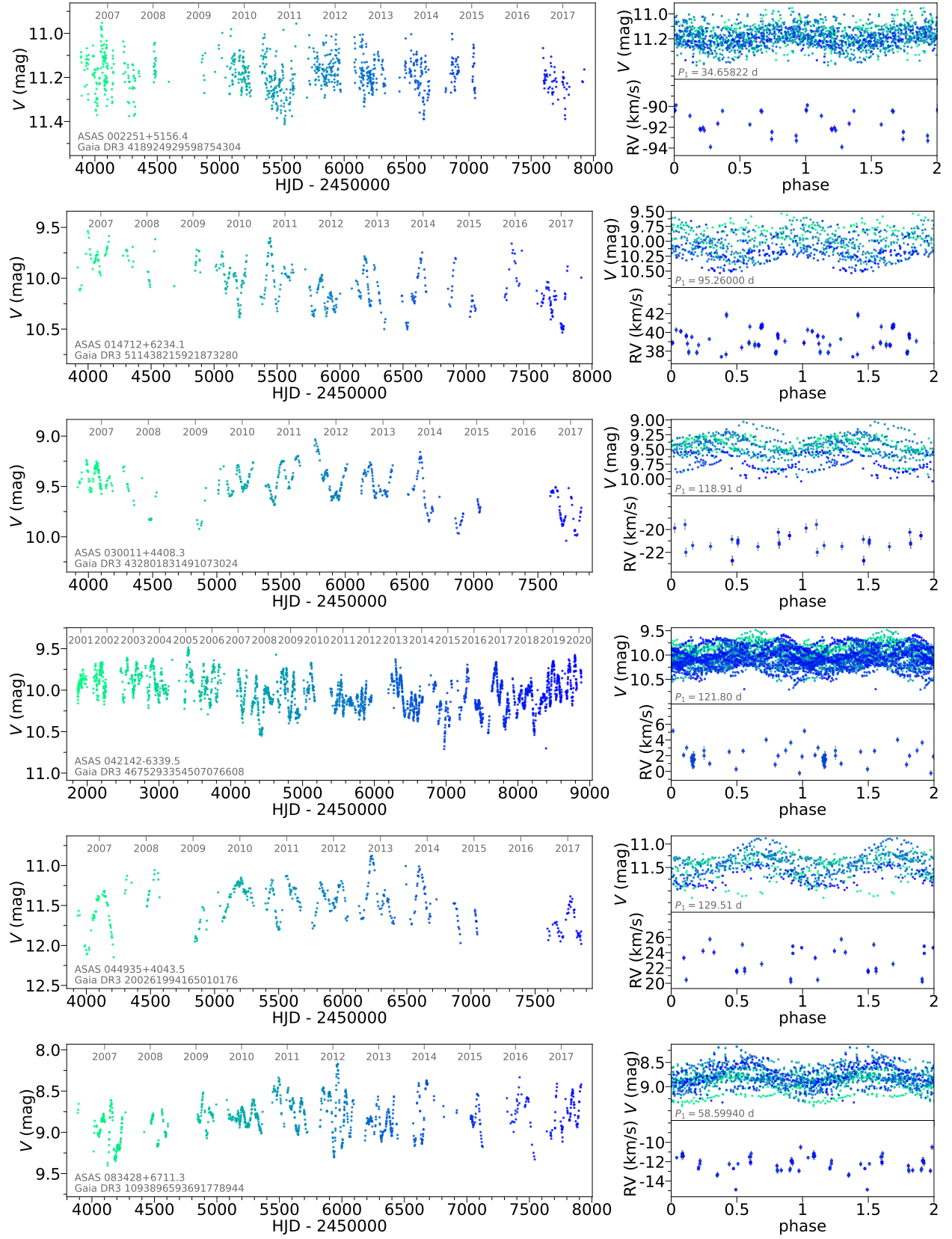

    \epsscale{1.1}
    \plotone{figures/SRV/002251+5156.4.pdf}
    \plotone{figures/SRV/014712+6234.1.pdf}
    \plotone{figures/SRV/030011+4408.3.pdf}
    \plotone{figures/SRV/042142-6339.5.pdf}
    \plotone{figures/SRV/044935+4043.5.pdf}
    \plotone{figures/SRV/083428+6711.3.pdf}
    \caption{Same as Figure \ref{fig:lsp1}, but for six stars from the S26 sample, which we reclassified as SRVs. The need for sufficiently long light curves for correct classification of LSP variability is evident.}
    \label{fig:srv1}
\end{figure*}

The remaining of the S26 sample is dominated by objects whose variability is more naturally interpreted as semi-regular rather than LSP-like. In total, 118 stars were classified as SRVs. Their ASAS light curves are characterized by irregular or multi-periodic behavior, changing amplitudes, and poor phase coherence on a long timescale. These properties are typical of evolved red giants with semi-regular pulsations, but they do not constitute convincing evidence for LSP. Consequently, these stars represent the main source of contamination in the Gaia FPR LSP sample. Six examples of such objects, treated as LSP by S26, but reclassified here as SRVs, are shown in Figure~\ref{fig:srv1}.

\section{\texttt{gaiamock} simulations}\label{sec:gaiamock-main}

In their analysis, S26 modeled the expected RUWE distribution for each of their LSP candidates, assuming that the Gaia FPR RV curves of these targets can be treated as a direct trace of Keplerian orbital motion. However, the authors' analysis did not allow for variation in the mass of the primary component (i.e., the giant star that dominates the total flux) in LSP systems, instead fixing it at $1\,{\rm M}_\odot$ level, while the population of LSP stars also includes intermediate-mass stars \citep[i.e., stars with initial masses $\gtrsim1.4\,{\rm M}_\odot$; e.g.,][]{saio2015, hofner2018, mcdonald2026}. Adopting only the lower mass limit for the primary component naturally leads to an overestimation of the expected RUWE value, due to the statistically larger size of the primary's absolute orbit. This, combined with the $53\%$ contamination of the sample with non-LSPs, has a significant impact on S26 results. Therefore, we decided to repeat the comparison between observed and expected RUWE values for a sample of confirmed LSP stars from the Gaia FPR catalog, as well as for a sample of the 23 brightest (and therefore closest to the Sun) LSP stars compiled by \cite{iwanek2025} based on the ASAS data. Two stars, namely EU\,Eri (ASAS ID 034419-4153.9, Gaia DR3 4848874011097672576) and RT\,Pav (ASAS ID 183630-6953.2, Gaia DR3 6431818980098135040), are included in both lists. Therefore, we removed these objects from the sample of \cite{iwanek2025} in our study to avoid duplication.

To enable a direct comparison of our simulated RUWE values for LSPs with those obtained by S26, we used the same version of the \texttt{gaiamock} Python package\footnote{Available for download on the \url{https://github.com/kareemelbadry/gaiamock} website.} \citep{el-badry-gaiamock, iorio-gaiamock-mod}. We used a so-called modified version of \texttt{gaiamock}, that is \texttt{gaiamock\_mod.py}, which is well-suited for predicting RUWE more reliably for binary systems with small photocenter orbits and for effectively single stars. This tool allows for estimating RUWE values for binary systems with pre-defined parameters. In particular, it allows determine whether fitting a 5-parameter astrometric model for a single star leads to an elevated RUWE value as a result of photocenter motion in an unresolved binary system. We performed our \texttt{gaiamock} simulations in two variants, an `agnostic' and an `educated' variant, which we describe in detail below.

\subsection{General framework and range of expected solutions}\label{sec:gaiamock-framework}
Estimating RUWE using \texttt{gaiamock} requires defining the following free parameters of a binary system: its total Gaia $G$-band magnitude, right ascension ($\alpha$) and declination ($\delta$), proper motion in right ascension multiplied by $\cos\delta$ ($\mu_{\alpha\ast}$) and in declination ($\mu_\delta$), true parallax\footnote{By `true parallax' we mean the inverse of the actual heliocentric distance to a binary system, not the measured parallax, which may systematically differ.} ($\varpi$), the masses of the primary ($M_1$) and secondary ($M_2$) components, orbital period ($P_{\rm orb}$), orbital eccentricity ($e$), time of periastron passage ($T_0$), argument of periastron ($\omega$), longitude of the ascending node ($\Omega$), orbital inclination ($i$), and the relative flux contrast of the secondary component with respect to the primary in the Gaia $G$ band ($F_{G,2}/F_{G,1}$). For the last parameter, we enforced $F_{G,2}/F_{G,1}=0$ in all of our experiments since the primary component is a giant star that outshines a much dimmer secondary component.

Before proceeding to the analysis of RUWE for specific LSP stars, we first examined the range of all possible relations between heliocentric distances ($d$) of LSPs and their RUWE (hereafter referred to simply as the `$d$\,--\,RUWE relation'). To achieve this goal, we randomly generated one thousand $d$\,--\,RUWE curves for various synthetic LSP systems. We assumed their random distribution across the sky, with their proper motions probed from the following realistic intervals $-15\leq\mu_{\alpha\ast}/({\rm mas}\,{\rm yr}^{-1})\leq 15$ and $-15\leq \mu_{\delta}/({\rm mas}\,{\rm yr}^{-1})\leq 15$. The $G$-band magnitudes were taken from the 8 to 18\,mag range, while the orbital periods covered the $200\leq P_{\rm orb}/{\rm d}\leq 1000$ range. To avoid imposing a rigid assumption on $M_1$, we allowed for primary masses from $1\,{\rm M}_\odot$ to $2\,{\rm M}_\odot$. For $M_2$, we covered the range of masses from $0.01\,{\rm M}_\odot$ to $0.25\,{\rm M}_\odot$. This interval corresponds to the most massive Jovian planets, through brown dwarfs, up to very low-mass stars. Following the results of \cite{nicholls2009}, we assumed orbital eccentricities in the range $0\leq e \leq 0.6$. We also assumed that all combinations of parameters determining the spatial orientation of the orbit (i.e., $\Omega$, $\omega$, $\cos i$) are equally probable. All LSP parameters required by \texttt{gaiamock} were drawn from independent uniform distributions.

\begin{figure*}
    \epsscale{1.2}
    \plotone{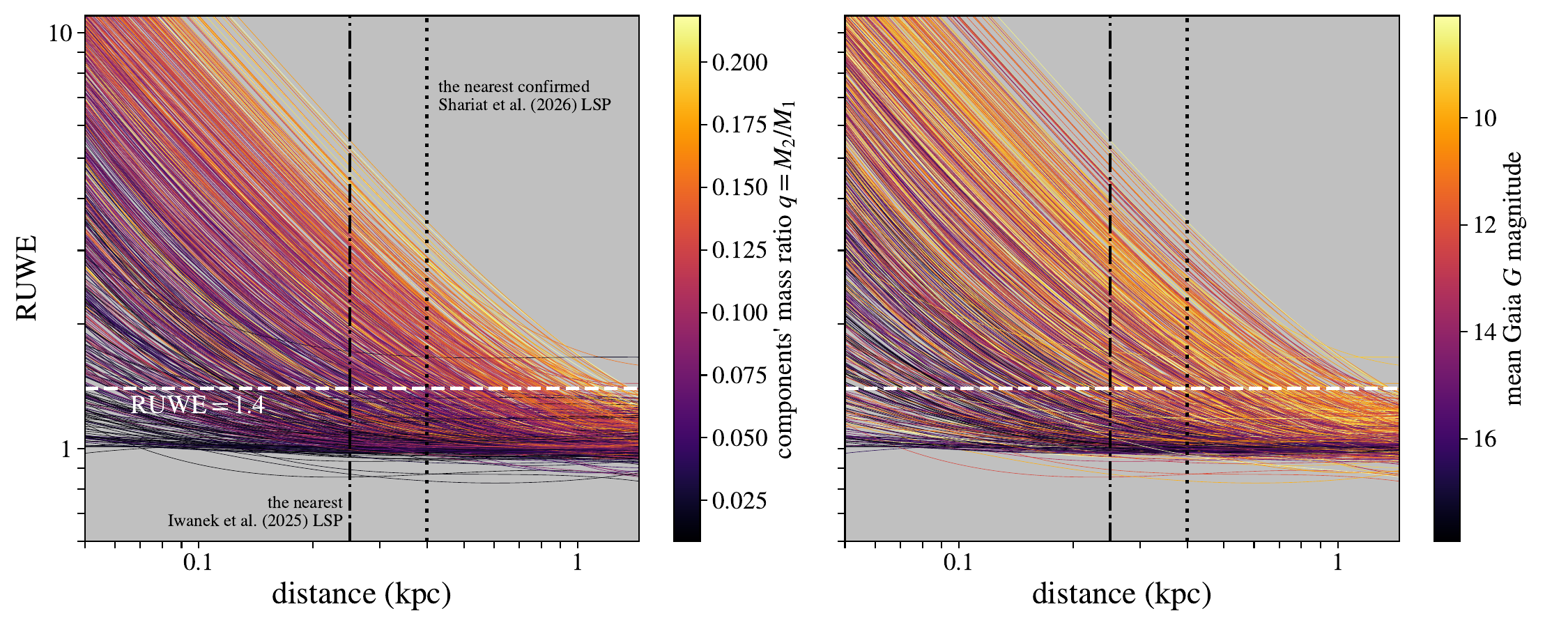}
    \caption{Simulated heliocentric distance\,--\,RUWE relation spectrum for synthetic LSP binary systems. Each individual curve represents one LSP system, whose RUWE evolves as its distance from the Sun varies. The curves are color-coded by the components' mass ratio (left panel) and the mean Gaia $G$ magnitude (right panel). In both panels, the horizontal white dashed line corresponds to ${\rm RUWE}=1.4$. The dash-dotted and dotted vertical lines indicate the distances to the nearest LSPs from the \cite{iwanek2025} sample and from the sample of LSPs confirmed by us in S26, respectively.}
    \label{fig:d-RUWE-curves}
\end{figure*}

The results of this experiment are summarized in Fig.~\ref{fig:d-RUWE-curves}. The $d$\,--\,RUWE relations exhibit a broad spread in terms of where they intersect the horizontal line ${\rm RUWE}=1.4$, which is commonly adopted as the threshold indicating an elevated RUWE value suggestive of binarity. We emphasize that for many potential LSP systems, their binary nature becomes discernible through RUWE only at distances of about 100\,pc or even smaller (curves close to the lower left envelope of $d$\,--\,RUWE distributions). Moreover, the distance at which the $d$\,--\,RUWE curve intersects the ${\rm RUWE}=1.4$ threshold is statistically smaller for systems with fainter mean $G$ magnitude and for those with smaller mass ratios, $q=M_2/M_1$. In Fig.~\ref{fig:d-RUWE-curves}, we also indicate the distances of the nearest LSP systems from our sample of confirmed S26 LSP stars, as well as from the sample delivered by \cite{iwanek2025}. This implies that both samples allow for the inference of binarity based on RUWE only for systems whose $d$\,--\,RUWE relations lie simultaneously above the white dashed horizontal line and to the right of the vertical lines. This, however, introduces an obvious observational selection effect, favoring the brightest LSP stars with relatively large mass ratios (i.e., secondaries preferably being low-mass stars). Indeed, only about $30\%$ of the whole spectrum of $d$\,--\,RUWE curves lies in the aforementioned privileged region\footnote{We emphasize that this does not mean that there is approximately $30\%$ chance of observing inflated RUWE, because in this section we only discuss the range spanned by the $d$\,--\,RUWE spectrum. The statistical probability that an LSP, being a binary system, would not exhibit inflated RUWE is discussed in Sects.~\ref{sec:agnostic-sims} and \ref{sec:educated-sims}.}. Conversely, if LSP systems are predominantly characterized by substellar companions (i.e., $M_2\lesssim0.08\,{\rm M}_\odot$, $q\lesssim0.06$), their binarity is almost undetectable via RUWE at the realistic distances of LSP stars.

\subsection{`Agnostic' simulations}\label{sec:agnostic-sims}
We decided to regenerate an equivalent of Fig.~4 from S26 for the LSP sample analyzed in our study. For each LSP star, we performed 10\,000 `agnostic' simulations to determine their individual distributions of observable RUWE in Gaia DR3. The statistical distributions of the free parameters in these simulations are summarized in the third column of Table~\ref{tab:gaiamock}. The agnosticism of these simulations is based on the fact that the possible binarity of LSPs is associated with numerous uncertainties regarding their orbital parameters and the mass of the secondary component. Consequently, we aim to impose the loosest possible constraints on individual parameters, assuming that the LSPs' RV curves cannot be directly modeled using a Keplerian orbit (more details are provided in Sect.~\ref{sec:educated-sims}).

Regarding heliocentric distances to LSP stars, we performed an analysis using both the distances originally reported by S26 and the geometric distances provided by \cite{bailer-jones2021}. Regardless of the LSP sample, their coordinates $\alpha$ and $\delta$, as well as their mean magnitudes in the Gaia $G$ band, were fixed and adopted from the Gaia DR3 catalog \citep{gaiadr3-mainsource2023}. We also treated the orbital period as a fixed variable, but its source varied for different groups of stars. For confirmed LSP variables from S26, we adopted $P_{\rm orb}=P_1$ from our independent analysis of ASAS data, while for the remaining objects in this sample that we reclassified as SRVs, we used the periods reported in the Gaia FPR. Next, for the sample of bright and nearby LSP stars of \cite{iwanek2025}, we directly adopted the $P_\mathrm{L}$ values from their Table~1 as $P_{\rm orb}$.

\begin{deluxetable*}{l c l l}
    \tablecaption{Statistical distributions of free parameters of LSP binary systems in our \texttt{gaiamock} simulations.\label{tab:gaiamock}}
    \tablehead{\colhead{Parameter} & Unit & \multicolumn{2}{c}{Prior distribution$^{({\rm a})}$}\\ 
    \colhead{} & \colhead{} & \colhead{`Agnostic' simulations} & \colhead{`Educated' simulations} }
    \startdata
    $\mu_{\alpha\ast}$ & (${\rm mas}\,{\rm yr}^{-1}$) & $\mathcal{N}(\texttt{pmra},\texttt{pmra\_error})$ & same\\
    $\mu_\delta$ & (${\rm mas}\,{\rm yr}^{-1}$) & $\mathcal{N}(\texttt{pmdec},\texttt{pmdec\_error})$ & same\\
    $\varpi$ & (mas) & $\mathcal{N}((d/{\rm kpc})^{-1},\texttt{parallax\_error})$ & same\\
    $M_1$ & $({\rm M}_\odot)$ & $\mathcal{U}(1,2)$ & same\\
    $M_2$ & $({\rm M}_\odot)$ & $\mathcal{U}_{\rm log}(0.01,0.25)$ & based on $f(M)$, $M_1$, and $\cos i$\\
    $T_0$ & (d) & $\mathcal{U}(0,P_1)$ & same\\
    $e$ & --- & $\mathcal{U}(0,0.6)$ & same\\
    $\Omega$ & (rad) & $\mathcal{U}(0,2\pi)$ & same\\
    $\omega$ & (rad) & $\mathcal{U}(0,2\pi)$ & $\mathcal{U}(\pi,2\pi)$\\
    $\cos i$ & --- & $\mathcal{U}(0,1)$ & $\mathcal{U}(0,\cos30^\circ)$\\
    $K_1$ & (${\rm km}\,{\rm s}^{-1}$) & --- & $\mathcal{N}(\texttt{amplitude\_rv},\texttt{amplitude\_rv}\times0.1)$\\
    \enddata
    \tablecomments{$^{({\rm a})}$ The distributions $\mathcal{N}(\mu,\sigma)$, $\mathcal{U}(x,y)$, and $\mathcal{U}_{\rm log}(x,y)$ denote a normal distribution with mean $\mu$ and standard deviation $\sigma$, a uniform distribution over the interval $[x,y]$, and a log-uniform distribution over the interval $[x,y]$ respectively. Variables given in monospace font refer to the original parameter names in the Gaia DR3 Main Source catalog, except \texttt{amplitude\_rv}, which are taken directly from Gaia FPR catalog of LPVs.}
\end{deluxetable*}

\begin{figure*}
    \epsscale{1.2}
    \plotone{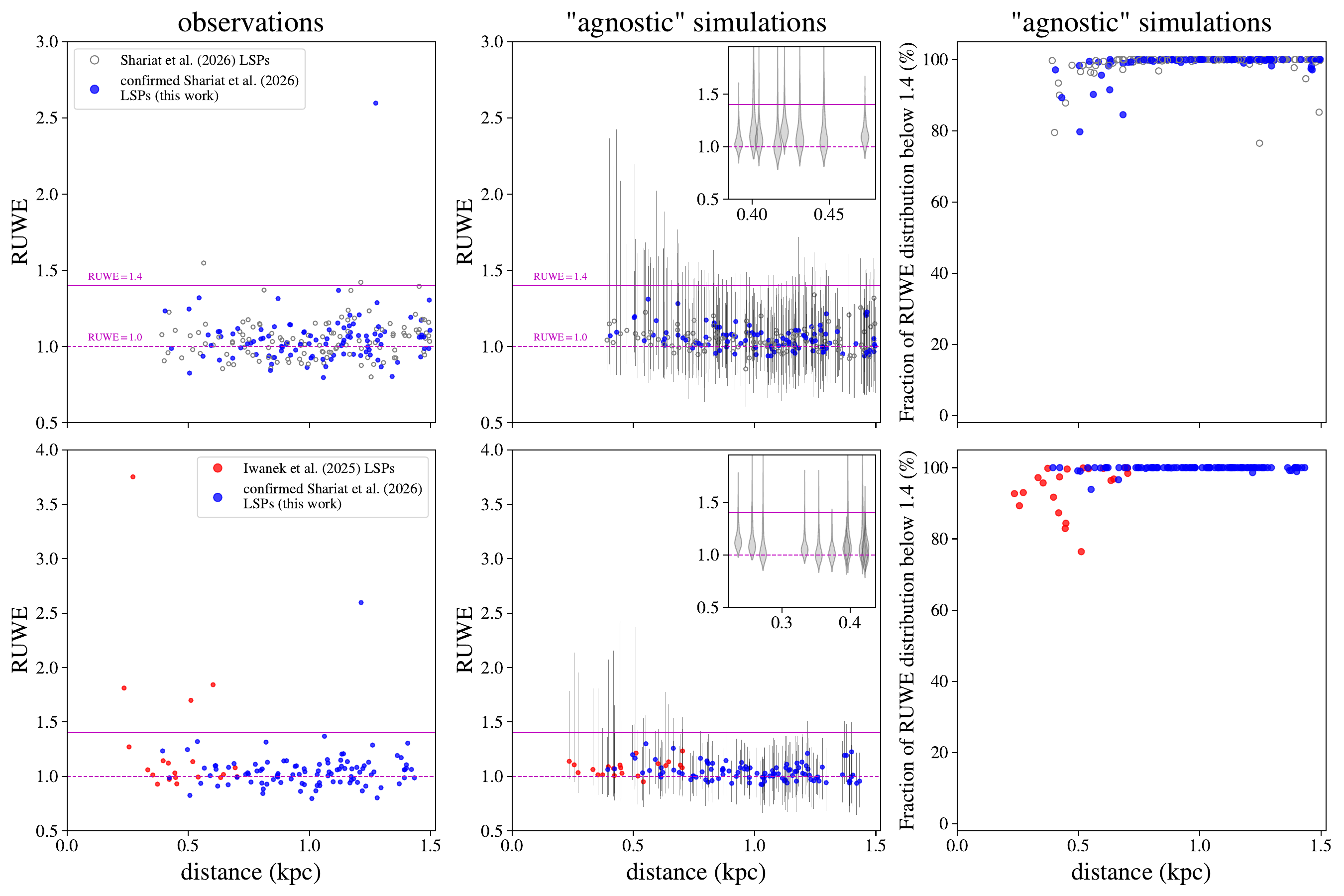}
    \caption{Comparison of the observed heliocentric distance\,--\,RUWE relations for different LSP samples with the results of our `agnostic' \texttt{gaiamock} simulations. \textit{Left column}: $d$\,--\,RUWE relations for the LSP samples indicated in the legends. The color coding of these samples is consistent across all panels in the figure. The horizontal solid and dashed magenta lines in the left and middle columns correspond to ${\rm RUWE}=1.4$ and $1.0$, respectively. \textit{Middle column}: expected RUWE distributions for each LSP system, estimated with \texttt{gaiamock}. The vertical gray lines mark the extrema of the theoretical RUWE distributions, while their medians are indicated by circles. The insets in the upper right corners show the full RUWE distributions for several of the nearest LSPs as violin plots. \textit{Right column}: percentage of each RUWE distribution lying below the ${\rm RUWE}=1.4$ threshold. In the top row, distances from S26 were adopted, while in the bottom row geometric distances from \cite{bailer-jones2021} were used. We note that the range of RUWE values on the ordinate axis is different in the upper and lower rows.}
    \label{fig:sims-agnostic}
\end{figure*}

The results of this analysis are presented in Fig.~\ref{fig:sims-agnostic}, where the observations are shown in the left column, while the simulations are in the middle and right columns. First, all individual RUWE distributions plotted in the middle panels exhibit a significantly skewed character, with most probability density concentrated around ${\rm RUWE}\approx1$ and a long tail extending towards higher RUWE values (see the insets). This behavior directly reflects the situation shown in Fig.~\ref{fig:d-RUWE-curves}, when taking a vertical slice at a fixed distance. Although the extremes of the RUWE distributions tend to broaden for closer LSPs, all of them are characterized by a median RUWE definitely below the $1.4$ threshold, and close to unity. It is striking that this remains true even for the nearest LSPs in the \cite{iwanek2025} sample (bottom middle panel). Second, and most importantly, each of the stars analyzed by us exhibits a RUWE distribution that extends below the ${\rm RUWE}=1.4$ level (right column). Indeed, even for the closest LSPs, the probability of displaying ${\rm RUWE}<1.4$ exceeds $\sim70\%$. This implies that for each of the LSPs, there is a genuine possibility of not exhibiting an elevated RUWE value despite their binarity and proximity to the Sun. In other words, although we assume all LSPs to be binary systems in our \texttt{gaiamock} simulations, this does not necessarily imply ${\rm RUWE\gtrsim 1.4}$, nor does it result in an obvious trend in the $d$\,--\,RUWE relation or in systematic vertical shift.

\begin{figure*}
    \epsscale{1.2}
    \plotone{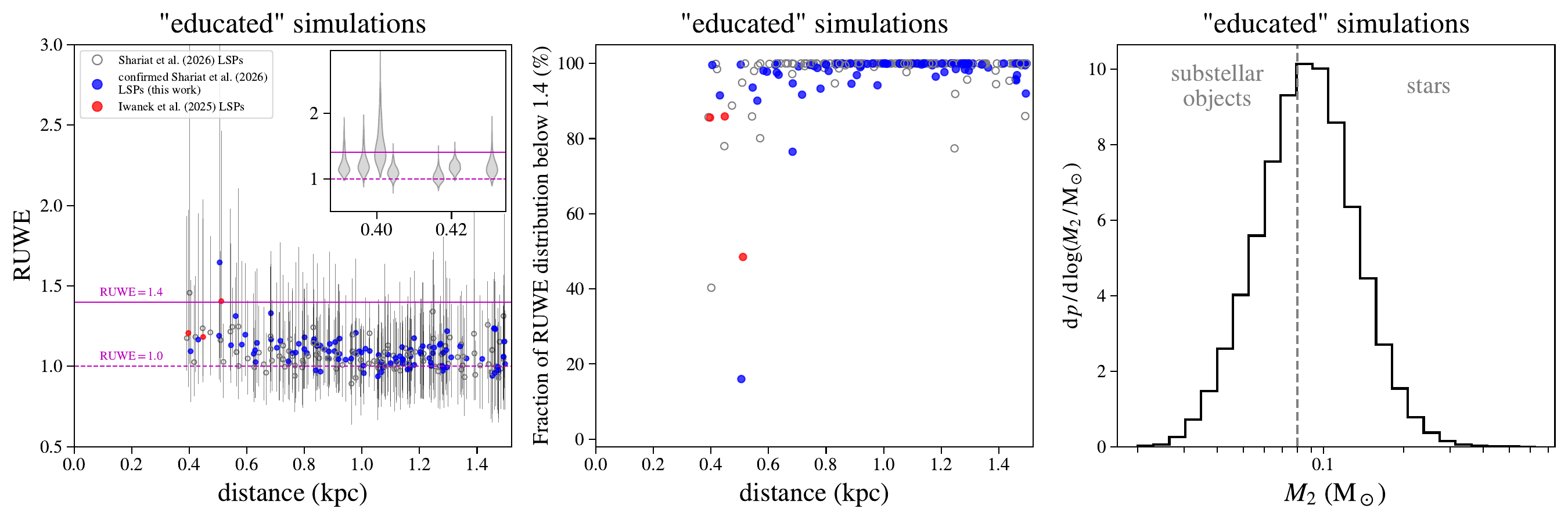}
    \caption{Summary of our `educated' \texttt{gaiamock} simulations. \textit{Left and middle panels}: same as in the upper middle and upper right panels of Fig.~\ref{fig:sims-agnostic}, but for the `educated' \texttt{gaiamock} simulations. \textit{Right panel}: Joint probability density distribution of RV-inferred companion masses for confirmed LSPs from S26 and \citet{iwanek2025}. The vertical dashed line marks $M_2=0.08\,{\rm M}_\odot$ -- the borderline between stars and substellar objects. More details can be found in the main text.}
    \label{fig:sims-educated}
\end{figure*}

\subsection{`Educated' simulations}\label{sec:educated-sims}
The `agnostic' simulations described above yield the broadest possible RUWE distributions for each LSP star, as they impose no constraints beyond those that are strictly necessary. However, for LSP stars from the Gaia FPR sample, RV curves are available, which can serve as additional constraints in our simulations. We therefore adopt a relatively strong assumption that the observed RV curves of LSP variables arise solely from the Keplerian motion of the primary's geometrical center on an eccentric orbit. We neglect the contribution of stellar pulsations to the overall surface velocity field, as well as non-Keplerian effects in the RV measurements arising from partial obscuration of the primary's disk by the expected dust structures. These `educated' simulations are carried out analogously to the `agnostic' ones, with several differences highlighted in the fourth column of Table~\ref{tab:gaiamock}. In this case, we utilize the RV semi-amplitude of the primary component ($K_1$) directly from the Gaia FPR. However, this catalog does not provide uncertainties for $K_1$. Based on Figs.~10 and 11 in S26, we adopt a realistic assumption in our `educated' simulations that this uncertainty amounts to $10\%$ of the $K_1$ value. Following \cite{decin2025} (their Sect.~2.3), we also impose a constraint on the minimum orbital inclination angle, which must now satisfy the relation $i>30^\circ$. This assumption is motivated by the requirement that the LSP system be sufficiently inclined toward the observer for the dust structure trailing the secondary component to obscure at least part of the primary's disk. By restricting the argument of periastron to the range $180^\circ<\omega<360^\circ$, we account for the well-established fact that LSP RV curves seem to be consistent with non-random orientation of Keplerian orbits \citep[][their Fig.~9]{nicholls2009}. Finally, $M_2$ is not treated as an independently sampled variable but is instead consistently computed from the mass function ($f(M)$) of a given LSP realization, its $M_1$, and $\cos i$. 

The effects of `educated' simulations are depicted in Fig.~\ref{fig:sims-educated}. We note that we were unable to perform the same analysis for all LSP stars from \cite{iwanek2025}, as only their five targets overlap with the Gaia FPR catalog of LPVs. Neglecting EU\,Eri and RT\,Pav, these additional objects are BM\,Eri, CI\,Phe, and DU\,Tuc. For this reason, only three LSP stars from \cite{iwanek2025} are shown in Fig.~\ref{fig:sims-educated}. The joint probability distribution of $M_2$ for genuine LSP stars, which is the result of the assumptions we described above, is shown in the right panel of Fig.~\ref{fig:sims-educated}. We note that it does not significantly differ from that presented by S26, except that the tail of the distribution towards higher masses is shorter in our case.

Surprisingly, the application of additional constraints related to $K_1$ and orbital geometry does not lead to any qualitative change in the conclusions drawn in Sect.~\ref{sec:agnostic-sims}. Nevertheless, it can be observed that the individual distributions of the expected RUWE become more symmetric compared to the `agnostic' case (see the inset in the left panel). The median RUWE values for the confirmed LSPs do not reveal any clear increase with decreasing distance (left panel). This is consistent with the fact that for the majority of LSPs, approximately $80\%$ of their RUWE distributions lie below the ${\rm RUWE}=1.4$ threshold. Similarly to the `agnostic' simulations, all confirmed LSPs are very likely to exhibit non-elevated RUWE values. It is highly suggestive that even the closest of the confirmed LSPs ($d\approx 240\,{\rm pc}$) is characterized by an almost $100\%$ probability of having ${\rm RUWE}<1.4$. Moreover, the majority of `educated' RUWE distributions still allow for individual RUWEs to be scattered around unity. In summary, if LSP variables are binary systems, detecting this feature through the analysis of RUWE is difficult in practice, given their distances.

\newpage
\section{Conclusions} \label{sec:conclusions}

We present an independent reassessment of the nearby Gaia FPR LSP sample analyzed by S26, based on long-term ASAS photometry and Gaia FPR RV data. Our classification shows that the original sample is significantly contaminated by non-LSP stars with only 103 out of 221 stars ($\sim 47\%$) exhibiting variability consistent with the LSP behavior. This highlights the importance of long-baseline photometric monitoring for reliable identification of LSP variability, particularly given the possibility of time-dependent changes in the observed photometric signatures. Unfortunately, the Gaia FPR LPV catalog suffers from relatively short observing baselines and a small number of epochs, which negatively impact the purity of the catalog's sample of LSP candidate stars. Therefore, population-level inferences drawn from the unfiltered sample, including those related to astrometric diagnostics, may be affected by significant contamination from SRV stars.

Using \texttt{gaiamock} simulations, we explore the expected Gaia RUWE behavior for binary systems that span a~broad range of primary-component masses, mass ratios, orbital configurations, and distances. We demonstrate that even if all LSP stars are assumed to be binary systems, many of them are not expected to exhibit elevated RUWE values at Gaia DR3 precision, particularly at distances of several hundred parsecs or more. The resulting $d$\,--\,RUWE relations are therefore intrinsically broad and do not imply a simple one-to-one correspondence between binarity and RUWE exceeding the commonly adopted threshold of 1.4.

Consistent with this, the observed $d$\,--\,RUWE relation for the confirmed nearby LSP stars does not show a significant trend with distance and remains largely below the ${\rm RUWE}=1.4$ threshold. This behavior is consistent with the range of outcomes predicted by our simulations and does not, by itself, provide evidence against a binary interpretation of LSP variability.

Although some individual objects exhibit observed RUWE values higher than those reproduced in our simplified simulations, such differences are not unexpected given additional observational and astrophysical effects not explicitly included in the model, such as variability-induced astrometric noise, crowding-related systematics, and chromatic effects in Gaia astrometry.

\section{Summary} \label{sec:summary}
We would like to summarize this paper briefly, in only four sentences: our goal was not to prove that every LSP star is a binary system. Instead, we show that the Gaia RUWE result based on a heavily contaminated sample cannot be generalized into a rejection of binarity as the origin of the LSP phenomenon, which we support with \texttt{gaiamock} simulations. While RUWE can, in principle, serve as a hint toward binarity, it cannot be used as a general veto against the binary explanation of the LSP phenomenon. Eppur binaria non \'e esclusa.

\acknowledgments
We are indebted to the OGLE collaboration for the use of facilities of the Warsaw telescope at Las Campanas Observatory, Chile, for their permanent support and maintenance of the ASAS instrumentation and to the Observatories of the Carnegie Institution of Washington for providing the excellent site for the observations. We would like to thank Mr. Wayne Rosing who has kindly allocated space and provided support for ASAS instruments at The Faulkes Telescope North site on Haleakala, Maui, Hawaii in the years 2006-2017. PI, PKS and DMS acknowledge support from the European Union (ERC, LSP-MIST, 101040160). Views and opinions expressed are however those of the authors only and do not necessarily reflect those of the European Union or the European Research Council. Neither the European Union nor the granting authority can be held responsible for them. This work has been funded by the National Science Centre, Poland, through grant 2022/45/B/ST9/00243 to IS.

\bibliography{paper}{}
\bibliographystyle{aasjournal}

\end{document}